\documentclass{webofc}
\usepackage[varg]{txfonts}   

\usepackage{amsmath}
\usepackage{amssymb}
\usepackage{graphicx}
\usepackage{subfigure}
\usepackage{babel}
\usepackage{enumerate}

\newcommand{\Zm}{Z$_{\text{max}}$ }
\newcommand{\zm}{Z$_{\text{max}}$}

\newcommand{\Sm}{$s_{\text{max}}$ }
\newcommand{\sm}{$s_{\text{max}}$}

\newcommand{\Mev}{MeV/A }
\newcommand{\mev}{MeV/A}

\newcommand{\Xesn}{$^{129}$\text{Xe}+$^{nat}$\text{Sn} }

\newcommand{\imgpath}{.}

\begin{document}

\woctitle{International Nuclear Physics Conference 2013}
\title{Pseudo-critical clusterization in nuclear multifragmentation}

\author{
D. Gruyer\inst{1}\fnsep\thanks{\email{diego.gruyer@ganil.fr}}  
\and J.D. Frankland\inst{1}  
\and R. Botet\inst{2}  
\and M. P\l{}oszajczak\inst{1} 
\and E. Bonnet\inst{1}  
\and A. Chbihi\inst{1}
\and P. Marini\inst{1} \\for the INDRA collaboration
}

\institute{
GANIL, CEA-DSM/CNRS-IN2P3, Bvd. Henri Becquerel, F-14076 Caen Cedex, France 
\and Laboratoire de Physique du Solide, Universit\'e de Paris-Sud, F-91405 Orsay, France
}

\abstract{
In this contribution we show that the biggest fragment charge distribution in central collisions
of \Xesn leading to multifragmentation is an admixture of two asymptotic
distributions observed for the lowest and highest bombarding energies.
The evolution of the relative weights of the two components with bombarding
energy is shown to be analogous to that observed as a function of
time for the largest cluster produced in irreversible aggregation
for a finite system. We infer
that the size distribution of the largest fragment in nuclear multifragmentation
is also characteristic of the time scale of the process, which is
largely determined by the onset of radial expansion in this energy
range.
}

\maketitle

Phase transitions play a central role in many fields of physics.
Indeed, they allow us to investigate the equation of state
and phase diagram of the system under study. The case of nuclear multifragmentation,
as observed in intermediate energy heavy-ion collisions \cite{Borderie2008Nuclear},
provides a unique opportunity to study not only thermodynamical properties
of nuclear matter, but also phase transitions in \emph{finite} systems.

\begin{figure}[th]
\begin{centering}
\subfigure[$N\sim10^{4}$ particles]{\includegraphics[width=0.32\linewidth]{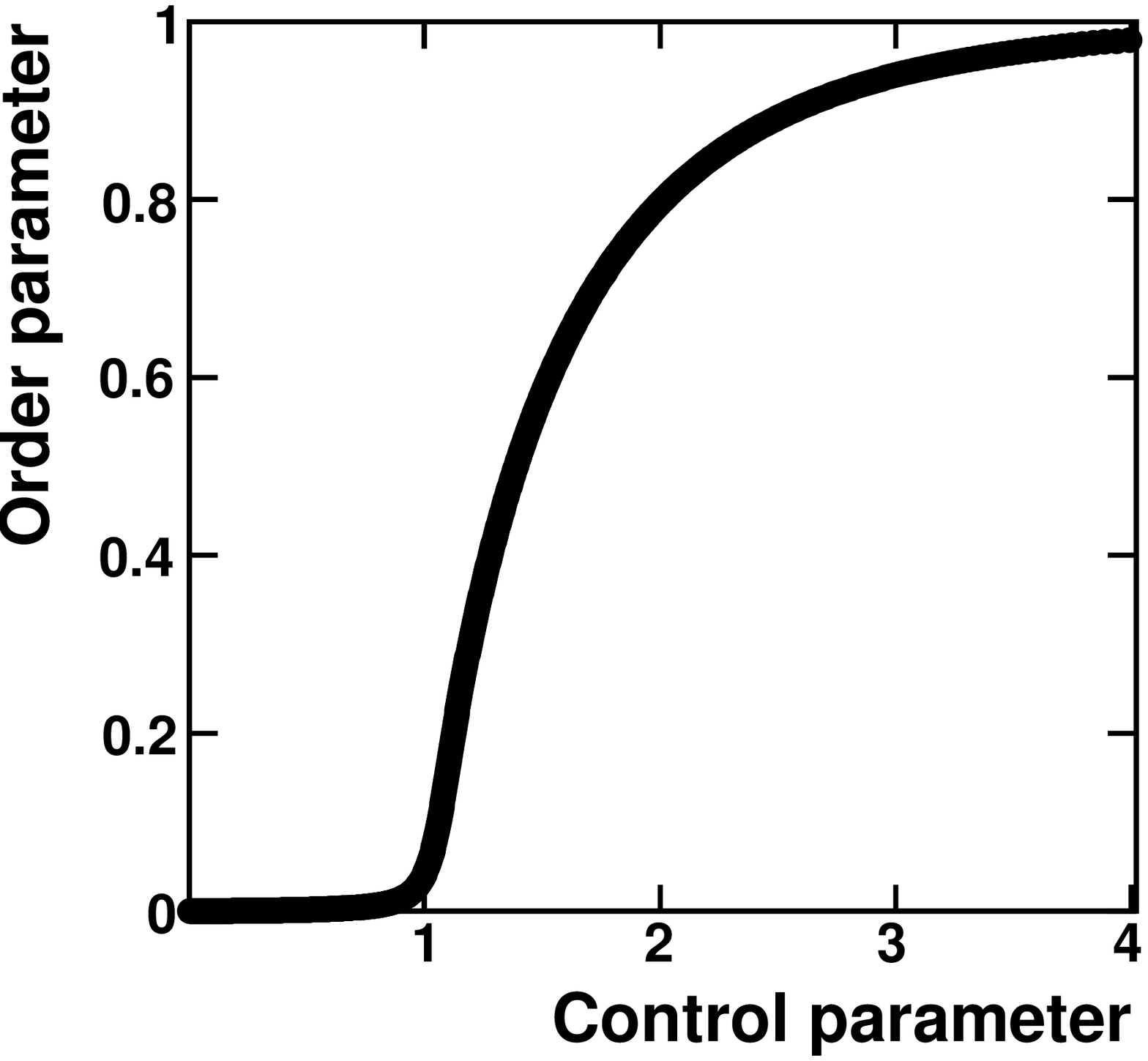}\label{ss10000}}\hfill
\subfigure[$N\sim200$ particles]{\includegraphics[width=0.32\linewidth]{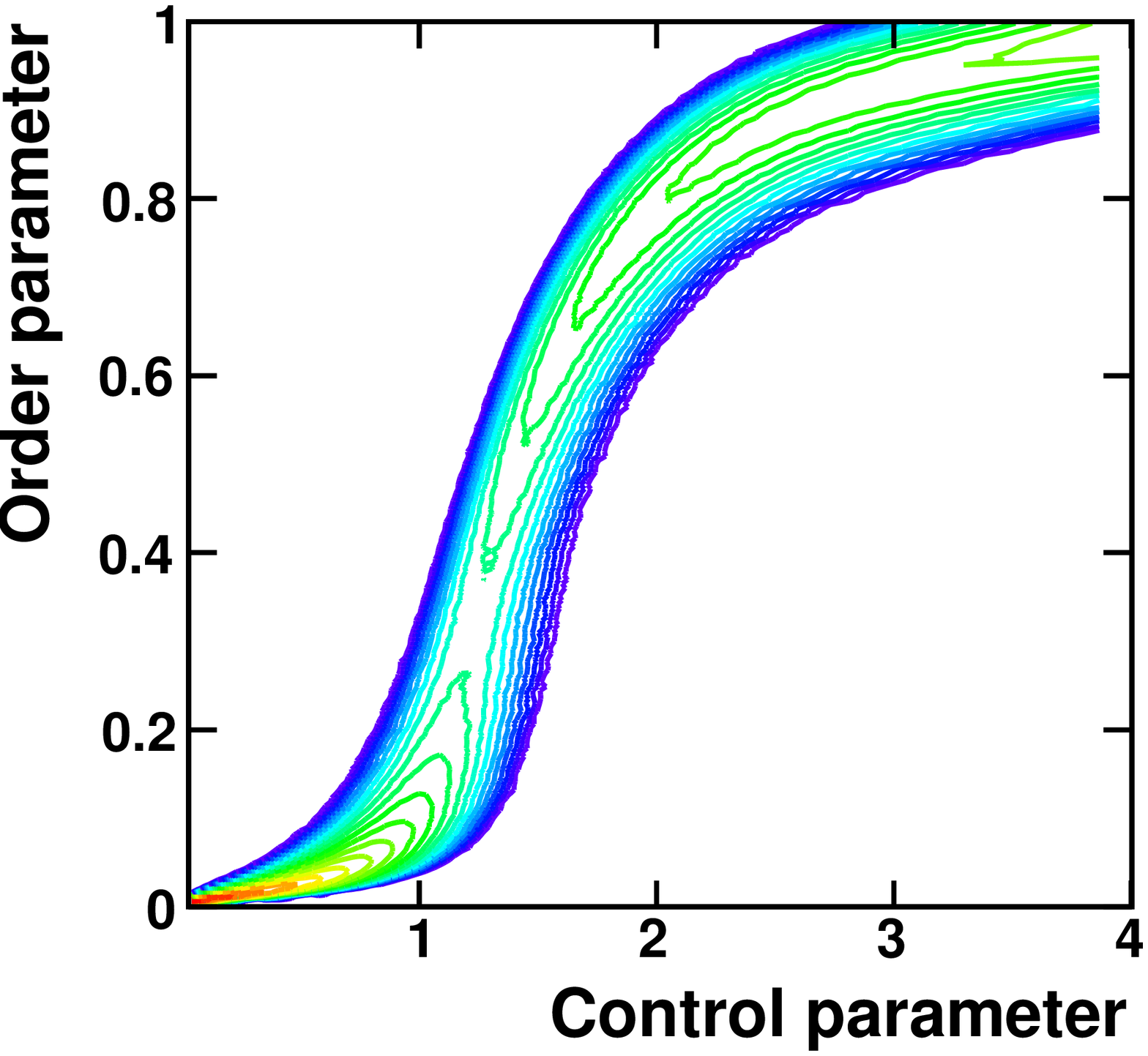}\label{ss216}}\hfill
\subfigure[$N\sim50$ particles]{\includegraphics[width=0.32\linewidth]{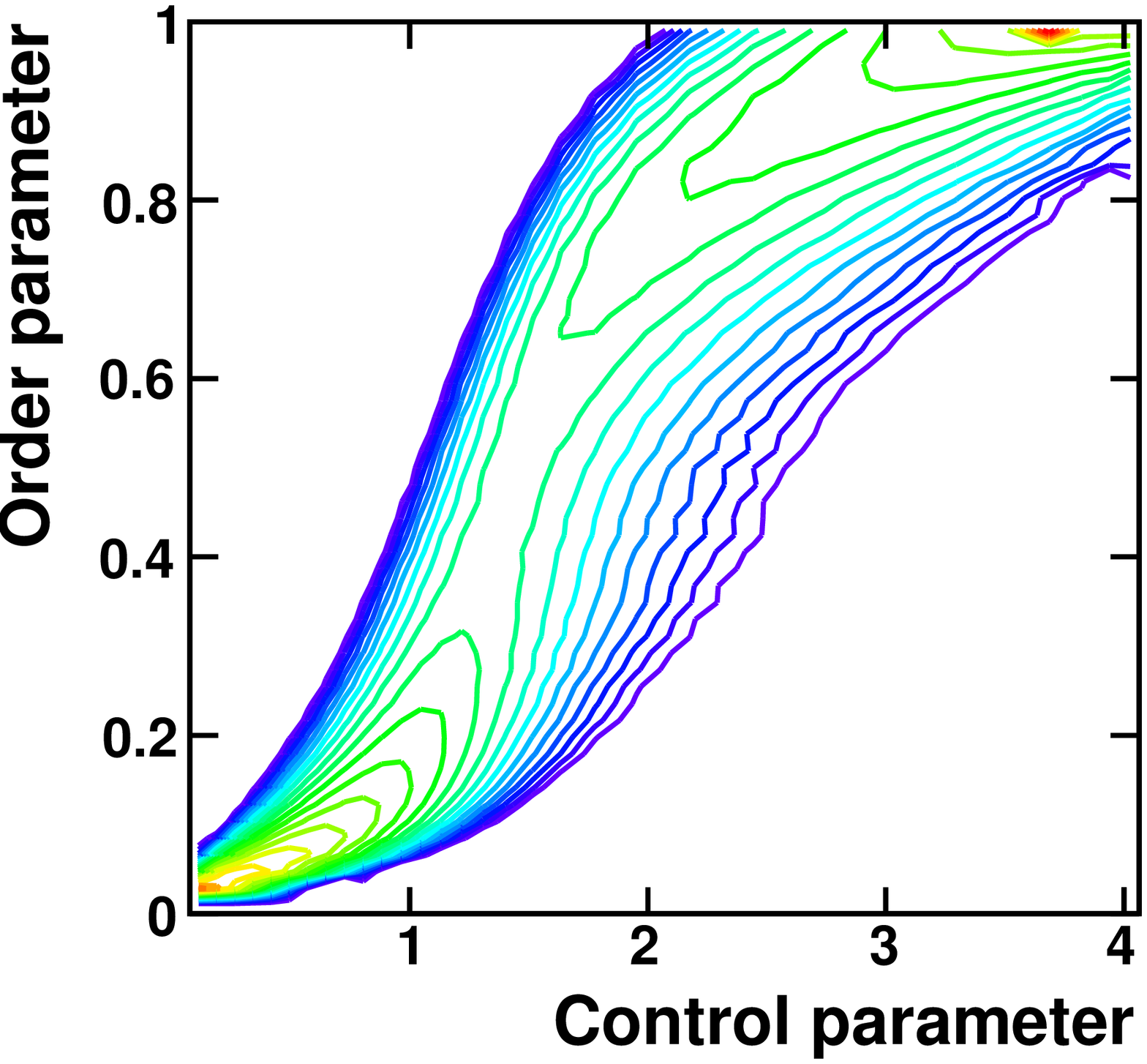}\label{ss54}}
\caption{(color online) Correlations between the order parameter and the control parameter for a critical phase
transition occurring in systems of different sizes.}
\end{centering}
\end{figure}

In an infinite system, fluctuations are generally irrelevant and phase 
transitions can be characterized by the correlations between the control parameter 
and the order parameter (fig. \ref{ss10000}). 
With decreasing size of the system, fluctuations become more important
and correlations are distorted (fig. \ref{ss216}), such that
for small system sizes (comparable to that of a nucleus), we are not able to identify
the transition from this simple picture (fig. \ref{ss54}).
The universal character of order parameter fluctuations in finite
systems \cite{Botet2000Universal} provides
a good framework in which to address such questions. Within
such a framework, it was shown that the size (atomic number)
of the largest fragment produced in multifragmentation events, \zm,
behaves like an order parameter, \emph{i.e.} the scaling properties
of its fluctuations change with increasing energy \cite{Botet2001Universal}.
In ref. \cite{Frankland2005Modelindependent} it was shown that a distinct
asymptotic form of the \Zm distribution 
can be associated with each scaling regime: a quasi-Gaussian 
distribution at the lower bombarding energies (E$_{\text{beam}} \lesssim 30$ \mev), and a Gumbel
distribution in the higher-energy disordered regime (E$_{\text{beam}} \gtrsim 40$ \mev). 

Here we study in more detail the transition from one energy regime
to the other, using new data on \Zm distributions for
\Xesn central collisions measured with INDRA \cite{Frankland2005Modelindependent,Bonnet2009Fragment},
at eight beam energies, between $25$ and $50$  \mev.
Within the hypothesis that at intermediate energies between the two regimes, \Zm
distributions can be described by an admixture of the two asymptotic forms \cite{Gruyer2013Nuclear}, 
we fit them with the following function :
\begin{eqnarray}
f(\text{\zm})=\eta f_{Ga}(\text{\zm})+(1-\eta)f_{Gu}(\text{\zm}),\label{fitEqua}
\end{eqnarray}
where, $f_{Ga}$ and $f_{Gu}$ are the Gaussian and the Gumbel component.

\begin{figure}[h]
\begin{centering}
\subfigure{\includegraphics[width=0.45\linewidth]{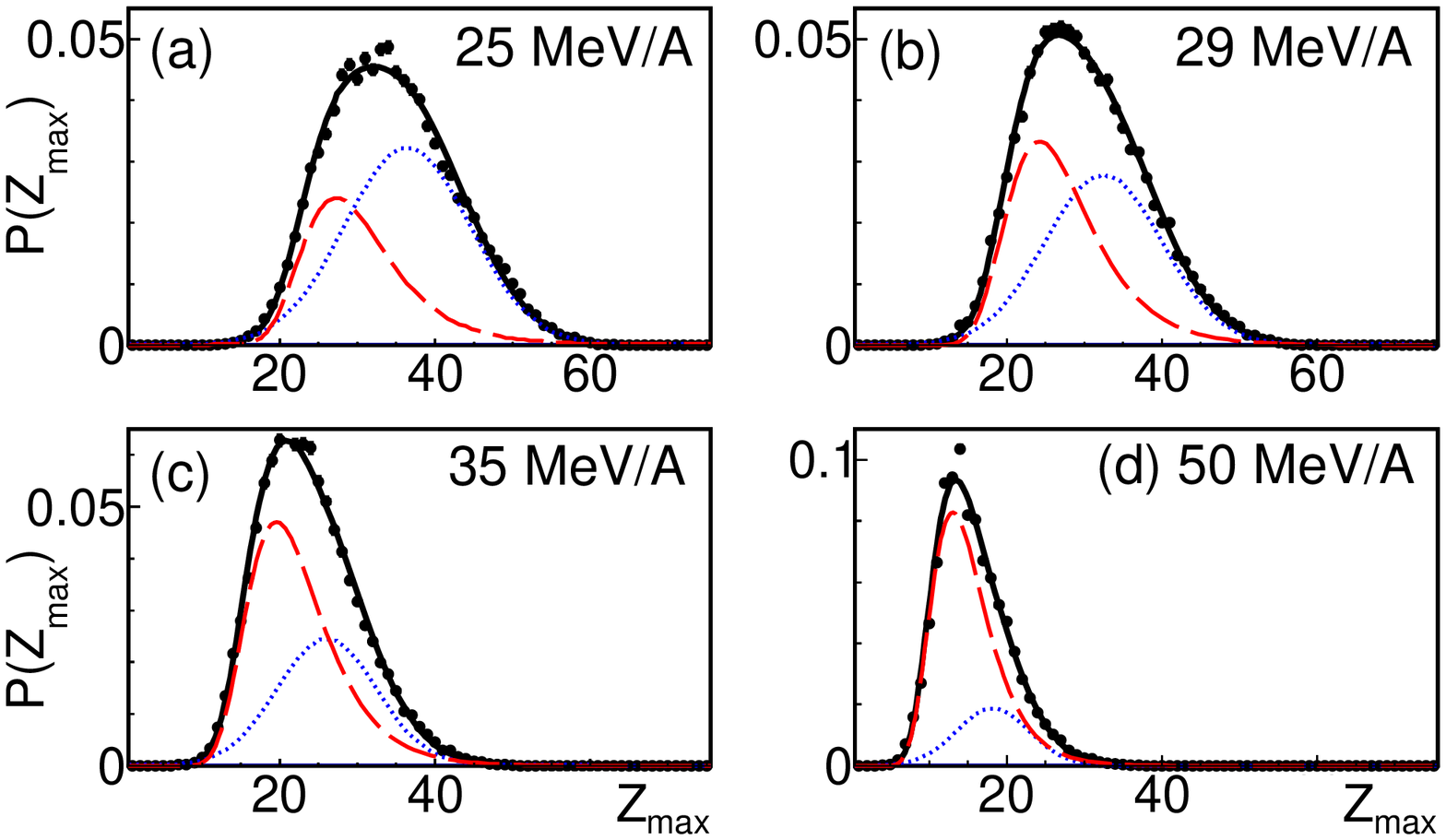}}\hfill
\subfigure{\includegraphics[width=0.45\linewidth]{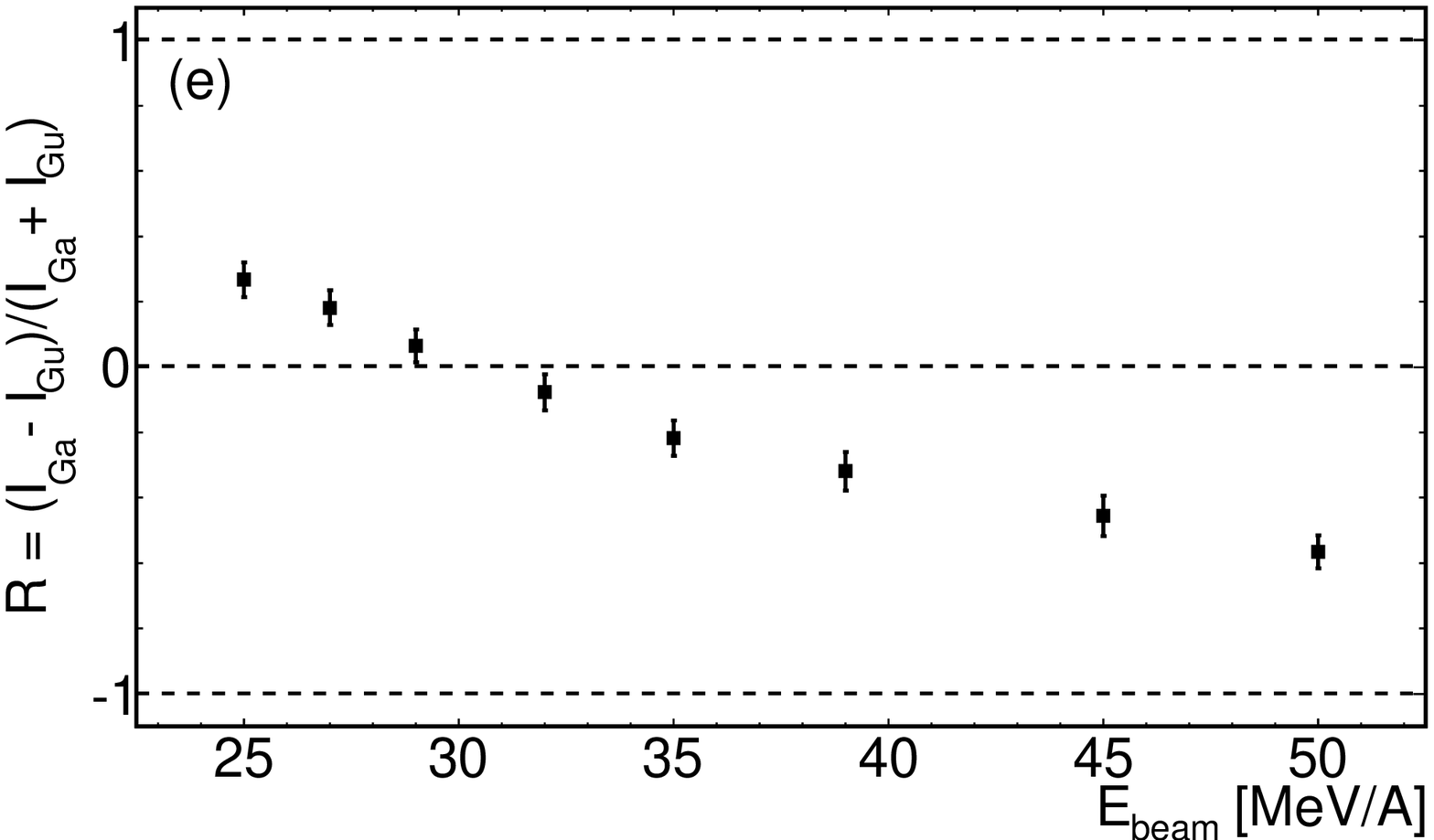}}
\caption{(Color online) Analysis of 
\Xesn experimental data: (a-d) largest fragment charge distributions, 
(Black solid curve) best fit to the data
using eq. (\ref{fitEqua}), (Red dashed curve) Gumbel component, (Blue dotted curve)
Gaussian component; and (e) evolution of the relative weight between the two
components, as a function of the beam energy.\label{fig:fitXeSn} }
\end{centering}
\end{figure}

The results of the fits using eq. (\ref{fitEqua})
are shown in fig. \ref{fig:fitXeSn}. It can be seen that, for all
analyzed energies, experimental data are well reproduced by $f(\text{\zm})$. 
At the highest considered energy (50 \mev), P(\zm)
is an almost pure Gumbel distribution (fig. \ref{fig:fitXeSn}(d)).
For lower bombarding energies (fig. \ref{fig:fitXeSn}(a-c)), both Gaussian and Gumbel 
contributions are present,
and the relative importance of the Gaussian component increases with
decreasing energy. In fig. \ref{fig:fitXeSn}(e) we present the evolution of
the relative weight between the two components as a function of the beam energy.
We observe not only two asymptotic regimes, but a continuous and smooth evolution between them.

To interpret such evolution, we consider the Smoluchowski irreversible aggregation model, 
which describes an out-of-equilibrium clusterization process. It exhibits a second-order 
phase transition after a critical time ($t_c$), whose order parameter
is also the average size of the largest cluster, \sm.
The order parameter distribution for irreversible aggregation
is not known exactly. 
For short time evolution ($t \ll t_c$), there are only few aggregation
events. So \Sm is the largest of randomly distributed cluster sizes.
\Sm is then of extremal nature, and it results in a Gumbel distribution (fig. 3(a)).
In this model, the average size of the largest cluster increases with
time as more and more coalescence of smaller clusters
takes place. At long times ($t \gg t_c$), the order parameter is then
essentially of additive nature. From the central limit theorem,
this results in an asymptotic Gaussian distribution (fig. 3(d)).
In the critical domain, \emph{finite size}
fluctuations are so large that similarly prepared systems
can exhibit one or the other behavior. In this domain, the \Sm distribution
is an admixture of the two asymptotic forms (fig. 3(b-c)), with a continuous
evolution of their relative weight (fig. 3(e)).

\begin{figure}[th]
\begin{centering}
\subfigure{\includegraphics[width=0.45\linewidth]{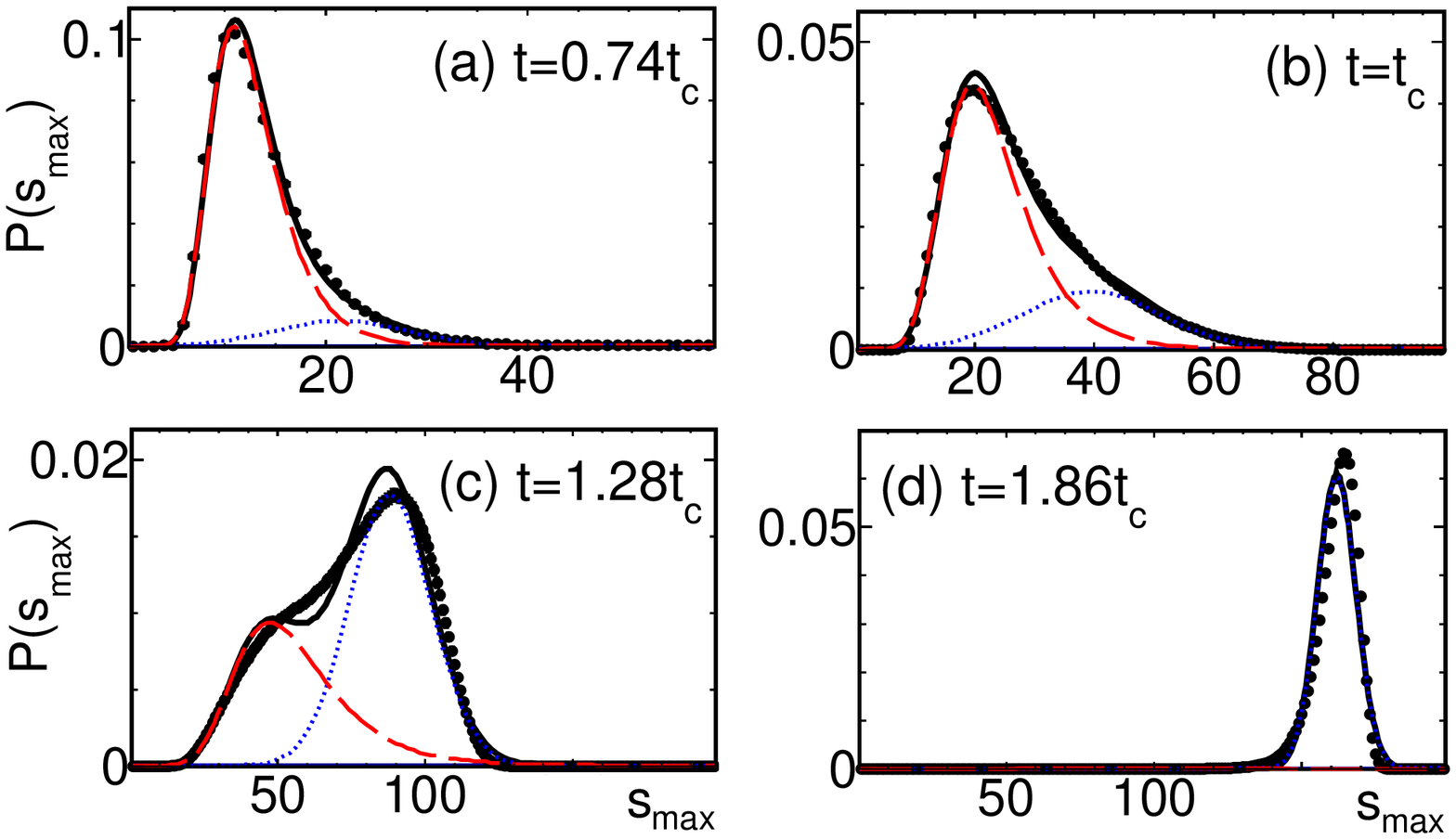}}\hfill
\subfigure{\includegraphics[width=0.45\linewidth]{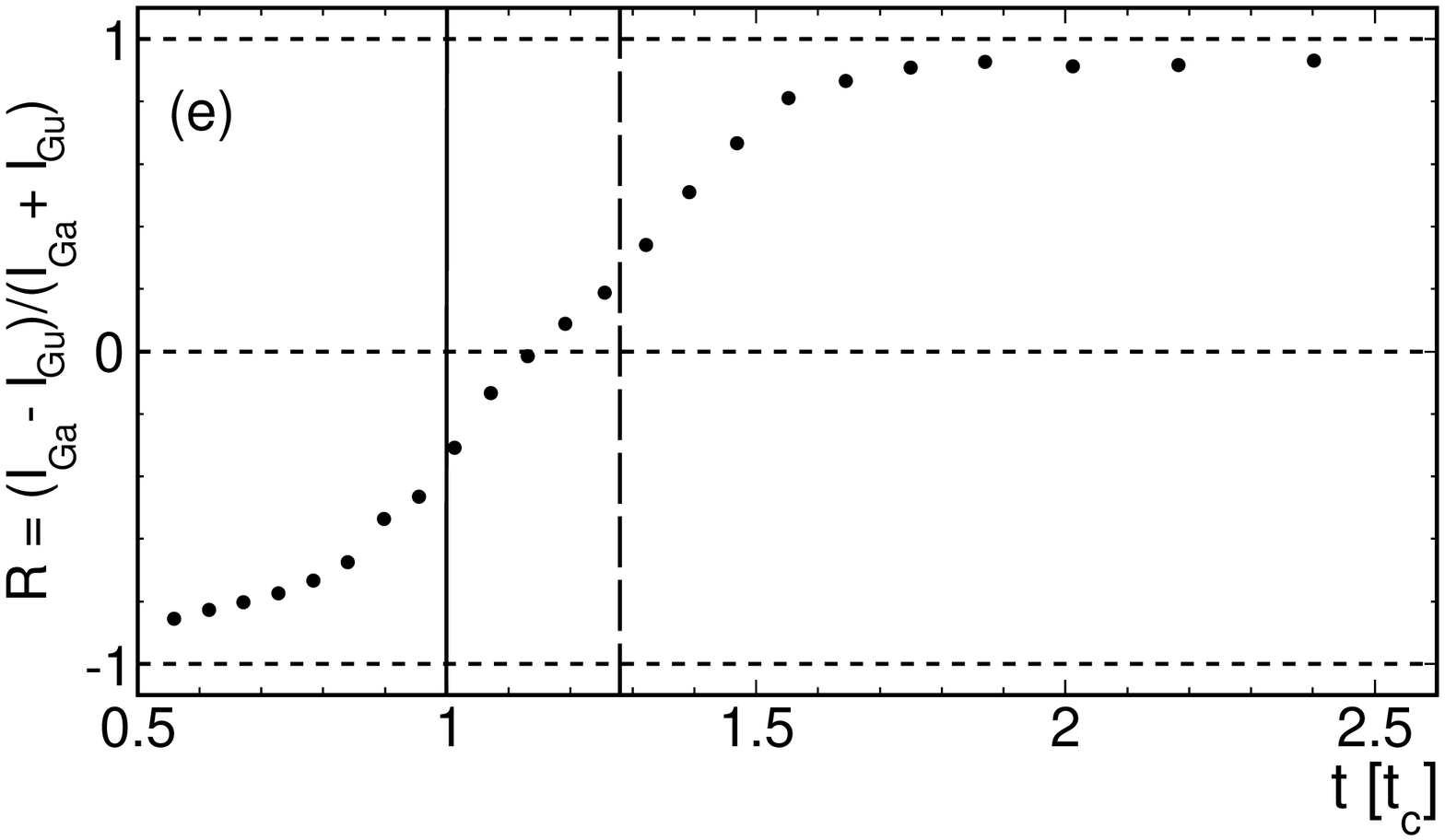}}
\caption{(Color online) Analysis of Smoluchowski calculations for $N=216$
particles: (a-d) largest cluster size distributions, (Black solid curve) best
fit to the data using eq. (2), (Red dashed curve) Gumbel component,
(Blue dotted curve) Gaussian component; (e) time evolution of the relative weight
between the two components (eq. (3)), 
(vertical dashed line) pseudo-critical time $t_{c}^{*}$.\label{fig:ResultSmolu} }
\end{centering}
\end{figure}

A strong similarity can be seen between the results of the analysis
of irreversible aggregation calculations (fig. \ref{fig:ResultSmolu}) 
and experimental data on \Xesn central
collisions (fig. \ref{fig:fitXeSn}). In the aggregation
model the order parameter distribution depends on the time
during which clusters can form. In central heavy-ion collisions, a
determining factor for the time-scale of fragment formation is the
radial expansion of the multifragmenting system, which increases with
the bombarding energy \cite{Borderie2008Nuclear}. Fragment sizes
evolve as long as exchanges of nucleons take place between them,
\emph{i.e.} until the freeze-out condition is reached. It has been
shown that, for central \Xesn reactions, the onset of significant radial
expansion occurs for beam energies above $25$ \Mev \cite{Bonnet2009Fragment}.
Therefore, the similarity between figs. \ref{fig:ResultSmolu} and \ref{fig:fitXeSn}
can be interpreted in terms of fragment size distributions
being determined on shorter and shorter time-scales due to increasingly
rapid expansion.

We have shown that the largest fragment size distribution
in multifragmentation events is an admixture of the two asymptotic
forms.  A similar decomposition is observed in critical aggregation, 
indicating that the critical domain lies around E$_{\text{beam}}\approx30$\Mev for
the \Xesn system. 
We interpret such criticality as the onset of an `explosive' multifragmentation regime.

\begin{acknowledgement}
The authors would like to thank the staff of the GANIL Accelerator
facility for their continued support during the experiments. D. G.
gratefully acknowledges the financial support of the Commissariat
\`a l'\'energie Atomique and the Conseil R\'egional de Basse-Normandie.
\end{acknowledgement}

\bibliography{articles}

\end{document}